\documentclass[11pt]{article}
\usepackage{latexsym}
\usepackage{graphicx}
\usepackage{amsfonts,amssymb}

\begin{document}

\title{The Community Structure of Econophysicist Collaboration Networks}
\author{Peng Zhang$^{1}$,Menghui Li$^{1}$, Jinshan Wu$^{2}$, Zengru Di$^{1}$,
Ying Fan$^{1}$\footnote{Author for correspondence: yfan@bnu.edu.cn}, \\
\\ 1. Department of Systems Science, School of Management,\\
Beijing Normal University, Beijing 100875, P.R.China
\\2. Department of Physics \& Astronomy, University of British Columbia, \\ Vancouver, B.C. Canada, V6T 1Z1}

\maketitle

\begin{abstract}
This paper uses a database of collaboration recording between
Econophysics Scientists to study the community structure of this
collaboration network, which with a single type of vertex and a
type of undirected, weighted edge. Hierarchical clustering and the
algorithm of Girvan and Newman are presented to analyze the data.
And it emphasizes the influence of the weight to results of
communities by comparing the different results obtained in
different weights. A function D is proposed to distinguish the
difference between above results. At last the paper also gives
explanation to the results and discussion about community
structure.
\end{abstract}



\section{Introduction}

In recent years, as more and more systems in many different fields
can be depicted as complex networks, the study of complex networks
has been gradually becoming an important issue. Examples include
the world wide web, social networks, biological networks, food
webs, biochemical networks and so
on\cite{Review1,Albert,SIAM,jazz,PNAS,Williams}. As one of the
important properties of networks, community structure attracts us
much attention. Community structure is the groups of network
vertices. Within the groups there have dense internal links among
the nodes, but between groups the nodes loosely connected to the
rest of the network\cite{Girvan}. Communities are very useful and
critical for us to understand the functional properties of complex
structure better. So the problem of detecting and analyzing
underlying communities is an important mission to us.

The idea of finding communities is closely related to the graph
partitioning in graph theory, computer science and
sociology\cite{PRE}. The study of community structure in networks
has a long history, so several types of algorithms have been
developed for finding the community structure in networks. Early
algorithms such as Spectral bisection\cite{EPJB} and the
Kernighan-Lin algorithm\cite{EPJB} perform poorly in many general
cases. To overcome the problems, in recent years, many new
algorithms have been
proposed\cite{Girvan,Wu,Reichardt,Fortunato,Donetti}. As one of
these algorithms, the algorithm of Girvan and Newman (GN) is the
most successful one. It is a divisive algorithm. The idea behind
it is edge betweenness, a generalization of the betweenness
firstly introduced by Freeman\cite{Freeman}. The betweenness of an
edge in network is defined to be the number of the shortest paths
passing through it. It is very clearly that edges which connect
communities, as all shortest paths connect nodes in different
communities have to run along it, have a larger betweenness value.
By removing the edge with the largest betweenness at each step, we
can gradually split the whole network into isolated components or
communities.

The application of GN algorithm has acquired successful results to
different kinds of networks\cite{Girvan,jazz}. Such as in
\cite{jazz}, the authors use the GN algorithm to study the
community structure of the collaboration network of jazz
musicians. The analysis to the results reveals the presence of
communities which have a strong correlation with the recording
location of the bands, and also shows the presence of racial
segregation between the musicians.

In recent years, most of the real-worlds which have been studied
were represented as non-weighted networks by neglecting lots of
data. The researchers paid more attention to the communities under
the influence of the topology of the network. However, the weight
of edges is important and may affect the results of communities,
and it can tell us more information than whether the edge is
present or not. For example, in a social network there are
stronger or poorer connections between individuals, and the weight
of edges are applied to describe the different strengths. So when
we try to detect communities in this network, we should consider
the weights into the process. It may give us better results
closely according with facts than ignoring them.

In \cite{Fan}, we built an Econophysics Scientific Collaboration
Network and gave some statistical results about this network. In
this paper, we focus on the investigation of community structure
of this network. We get the results of communities by using GN
algorithm and hierarchical clustering. We also obtained the
communities in different conditions including weighted,
non-weighted, and different weights. In latest months, Newman has
pointed out that applying the original GN algorithm to the
weighted networks would obtain poor results, and gave the
generalization of the GN algorithm to a weighted
network\cite{Weighted}. In \cite{PRE}, Newman and Girvan define a
function Q to measure where the best division for a given network,
and also a generalization of Q to the weighted networks was
proposed in this paper. We applied it to our network and found the
peaks of Q correspond closely to the expected divisions.

The outline of this article is as follows. In Section \ref{model}
we describe in detail our work. First, we introduce our database
and the Econophysics Scientific Collaboration Network which was
built on the database briefly. Then we describe the algorithms and
the definition of the different weights that were used to find
underlying communities in our network. At last, we give the
results by each condition and the compare between them. In Section
\ref{conclud} we give our conclusions.

\section{The communities acquired by different algorithms}\label{model}
A network is composed of a set of vertices and edges which
represent the relationship between the two nodes. In the
Econophysicist collaboration network\cite{Fan}, each node
represents one scientist. If the two scientists have collaborated
one or more papers, they would be connected by an edge. In order
to distinguish the different level of collaboration, we define the
weights on the edges. So it's a weighted network. Here we take the
largest cluster from the network as the subject of our research.
It is a sparse network including 271 nodes and 371 edges.

The weight is the crucial factor in our network analysis. Edge
weights represent the strength or capacity of the edges. The
weight of this network is defined as: $w_{ij}=tanh(t_{ij})$, where
$t_{ij}$ is the number of papers which the researchers have
collaborated. The reason we prefer the $\tanh$ function in
empirical studies is that, first, it has the saturation effect,
which makes the contribution less for larger connecting times;
second, it normalizes the maximum value to $1$, which is the usual
strength of edge in non-weight networks\cite{Fan}. As the
similarity is used here as the weight, the larger the weight is,
the closer the relation between the two ends nodes is. The weight
and connection provide us a natural description for the distance
of two nodes.

In this part, we present two methods, hierarchical clustering and
the algorithm of GN, on the analysis of community structure in our
network. Because GN algorithm performs well in many networks and
hierarchical clustering is the principal technique used in social
networks in current.

In practical situation the algorithms will normally be used on
networks for which the communities are not known ahead of time.
This raises a new problem: how do we know when the communities
found by the algorithm are good ones? To answer this question, in
\cite{Mixing}, Newman proposed a measure of the quality of a
particular division of a network, which they call it the
modularity. Then they define a modularity measure by
\begin{equation}
Q=\frac{1}{2m}\sum_{ij}[A_{ij}-\frac{k_ik_j}{2m}]\delta(c_i,c_j)
\label{Q}
\end{equation}

This quantity measures the fraction of the edges in the network
that connect nodes of the same community minus the expected value
of the same quantity in a network with the same community division
but random connections between the vertices. $A_{ij}$ represents
the edge between nodes $i$ and $j$, the degree $k_i$ is defined as
$k_i=\sum_{j}A_{ij}$ , and $m=\frac{1}{2}\sum_{ij}A_{ij}$. $c_i$
is the community to which vertex $i$ is assigned. Newman has
generalized the above measure to weighted networks\cite{Weighted}.
Here we use the similar formula to our weighted network with
similarity weight range from 0 to 1:
\begin{equation}
Q=\frac{1}{2m}\sum_{ij}[w_{ij}-\frac{w_iw_j}{2m}]\delta(c_i,c_j)
\label{Q}
\end{equation}
where $w_{ij}$ represents the weight in the edge between nodes $i$
and $j$, $w_i$ is the weight of node $i$: $w_i=\sum_{j}w_{ij}$ ,
and $m=\frac{1}{2}\sum_{ij}w_{ij}$.

Using hierarchical clustering method to find communities, we start
from an empty graph with all nodes and no edges. Then we connect
the edges in order of decreasing similarity. In our network, we
use the measure $d_{ij}$ which describes the similarity to be the
short path between a pair $(i,j)$ of vertices, where the shorter
path represents the bigger similarity. When the nodes are
clustered to be the communities, we define the distance between
different communities as follows
\begin{equation}
D_{pq}=\max_{i\in p,j\in q}d_{ij}.
\label{max}
\end{equation}

$p,q$ are any two communities. The measure $d_{ij}$ equals to the
shortest path between a pair $(i,j)$ of vertices.

We have got the result from above hierarchical clustering method.
It shows the modules in this result and also a peak in $Q$
function. The best division has 23 clusters.

GN method has got better results for community analysis. As
mentioned in the section of introduction, the idea behind the
algorithm of GN is edge betweenness. And the betweenness of an
edge is defined to be the number of the shortest paths passing
through it. To search the shortest paths between any two vertices,
we use the Dijkstra algorithm. For the determination of shortest
path, the similarity weight $w_{ij}\in\left[0,1\right]$ has been
transformed to dissimilarity weight by
$\tilde{w}_{ij}=\frac{1}{w_{ij}}$, and then
$\tilde{w}_{ij}\in\left[1,\infty\right]$ is corresponding to the
"distance" between nodes. All paths are calculated under this
dissimilarity weight from now on if not mentioned. The principal
ways of GN algorithm are as follows \cite{PRE}:

1.Calculate betweenness scores for all edges in the network.

2.Find the edge with the highest score and remove it from the
network.

3.Recalculate betweenness for all remaining edges.

4.Repeat from step 2 until all links are removed.

\begin{figure}
\center \includegraphics[width=12cm]{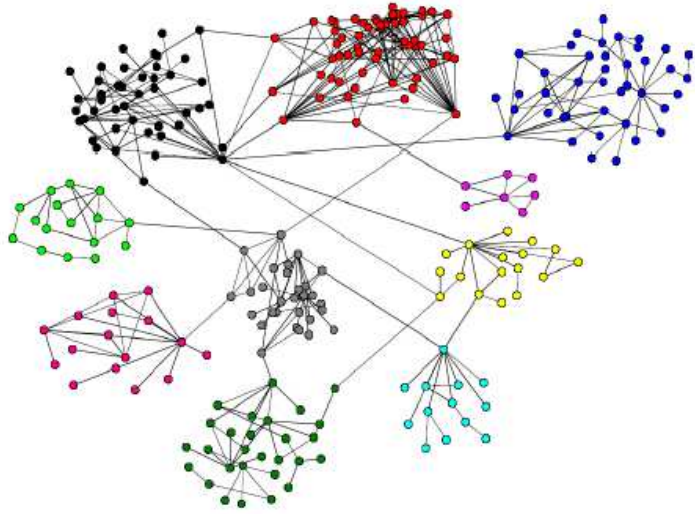} \caption{The best
division of Econophysics Scientist Collaboration Network, with the
divisions detected by GN algorithm represented by different
colors.}\label{a}
\end{figure}

The best result given by maximum $Q$ has 10 clusters. The GN
algorithm and the hierarchical clustering which are based on the
equation \ref{max} all show the modules in the results. In the
best divisions, we analyze the communities with the data. The
results of algorithm of GN is better than hierarchical clustering.
Because the result in the best division of GN algorithm shows that
the scientists, who are in the same university, institute or
interested in similar research topic, are clustered to one
community. It is close to the reality. For example, in figure
\ref{a} the members of the red community are most from Boston
University USA. And there are other communities which the members
are focused on the same topic, as the yellow one. Meanwhile even
the hierarchical clustering shows the modules, the result is not
consistent with the reality.


\begin{figure}
\center \includegraphics[width=7cm]{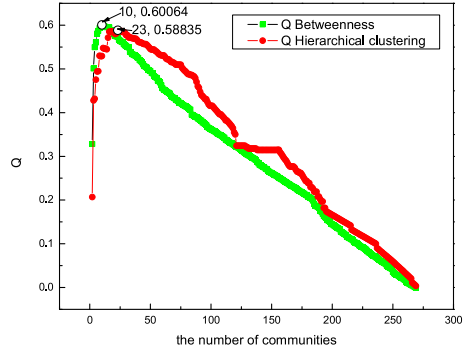}
\caption{The comparison of modularity detected by GN algorithm and
hierarchical clustering with Maximum minimum method. The peak are
in 10 and 23 clusters respectively.}\label{c}
\end{figure}

\section{The comparison of different formation of communities}
\label{analytic}

In the above section, we obtained that the results of GN algorithm
and hierarchical clustering are different. How to quantify the
difference between them? We define a function $D$ to measure it.
The idea behind the function is to discuss the similarity and
dissimilarity between sets $A$ and $B$. Let's discuss the
similarity and dissimilarity of two sets $A$ and $B$ defined as
subset of $\Omega$. The idea is quite trivial, the similarity is
represented by $A\cap B$, the dissimilarity should corresponds to
$\left(A\cap\bar{B}\right)\cup\left(\bar{A}\cap B\right)$.
Therefore, the normalized similarity and dissimilarity can be
defined as
\begin{equation}
\left\{
\begin{array}{l}
s = \frac{\left|A\cap B\right|}{\left|A\cup B\right|}\\
\\
d = \frac{\left|\left(A\cap\bar{B}\right)\cup\left(\bar{A}\cap
B\right)\right|}{\left|A\cup B\right|}
\end{array}\right. .
\end{equation}
In a way more convenient to be generalized to classification
systems with more than only two sets, we can rewrite above
expression by the characteristic mapping of set
$sign\left(X,\omega\right)$, which is defined as following,
\begin{equation}
sign\left(X,\omega\right) = \left\{
\begin{array}{ll}
1 & \mbox{for } \omega \in X
\\
\\ 0 & \mbox{for } \omega \bar{\in} X

\end{array}\right. .
\end{equation}
This mapping from $X$ to $\left\{1,0\right\}$ can be very
machinery calculated for any element $\omega$ in $\Omega$ and for
any subset $X$. It's easy to check
\begin{equation}
\left\{
\begin{array}{l}
\left|A\cap B\right| = \sum_{\omega\in\Omega}
sign\left(A,\omega\right)sign\left(B,\omega\right)
\\
\\
\left|\left(A\cap\bar{B}\right)\cup\left(\bar{A}\cap
B\right)\right| = \sum_{\omega\in\Omega}
\left|sign\left(A,\omega\right) - sign\left(B,\omega\right)\right|
\end{array}\right. .
\end{equation}
And also
\begin{equation}
\left|A\cup B\right| = \left|A\cap B\right| +
\left|\left(A\cap\bar{B}\right)\cup\left(\bar{A}\cap
B\right)\right|
\end{equation}
Therefore, by the characteristic mapping, the similarity and
dissimilarity are reexpressed by
\begin{equation}
\left\{
\begin{array}{l}
s = \frac{\sum_{\omega\in\Omega}
sign\left(A,\omega\right)sign\left(B,\omega\right)}{\sum_{\omega\in\Omega}\left[
sign\left(A,\omega\right)sign\left(B,\omega\right) +
\left|sign\left(A,\omega\right) -
sign\left(B,\omega\right)\right|\right]}
\\
\\
d =\frac{\sum_{\omega\in\Omega} \left|sign\left(A,x\right) -
sign\left(B,\omega\right)\right|}{\sum_{\omega\in\Omega}\left[
sign\left(A,\omega\right)sign\left(B,\omega\right) +
\left|sign\left(A,\omega\right) -
sign\left(B,\omega\right)\right|\right]}
\end{array}\right. .
\end{equation}

Consider a particular division of a network into $k$ communities.
There are two formations of $k$ communities by different
algorithms. we can deduce the comparison of them into many pairs
of comparison between sets.The principal way is:

1. Construct the correspondence between the two subsets from
different conditions

2. Compare every corresponding pair.

3. At last, integrate all the results from comparison of every
single pair.

Correspondence relation here means, for every subset $X_{i}$ in
classification $X_{k}$, to find the counterpart $Y_{i}$ in
$Y_{k}$, by the similarity measurement. Here $X_{i}$ and $Y_{i}$
correspond to above $A$ and $B$. After that we will get two
ordered set $X$ and $Y$, where the elements at the corresponding
order are a pair of counterparts. And then apply the dissimilarity
measure onto every pair to get a measurement of the total
dissimilarity.

\begin{equation}
D=\frac{\sum_{i=1}^{k}{d_{X_{i}Y_{i}}}}{k}
\end{equation}

Under this definition, $d$ will be normalized in
$\left[0,1\right]$, where $\left(0,1\right)$ means no and large
difference respectively.

The principle of the first step is to compare every single set
$X_{i}$ from $X$ with all the $Y_{i}$ in $Y$, and group it with
the one having largest similarity. However, at some cases, this
may lead to a very ugly correspondence, for instance, many $X_{i}$
correspond to the same $Y_{i}$. at this time, we choose the
largest one of them and group the $X_{i}$ which correspond the
largest one with the $Y_{i}$. other $X_{i}$ should found the
counterpart again in rest $Y_{i}$. But in some times, we want to
discuss the different formations of communities, for example, we
want to compare the dissimilar between the best division of GN
algorithm and hierarchical clustering. As obtained above, we know
that the number of communities are different. it meant some
$Y_{j}$ in hierarchical clustering  don't have any counterparts.
In this case, the first step still can be done by treating the
whole group as a large subset, and treating no counterpart as
empty set $\Phi$. The $k$ equal to the larger number. Here we give
two examples, the first one, a network, including $N$ nodes, was
divided into two communities by two algorithms. One division is
two equal communities. The other division is a node and the rest
nodes. Calculating the dissimilar of two algorithm, we got
$d\approx0.75$. The second example is a network, including $N$
nodes, was divided into $N$ communities. calculating the
dissimilar of the whole network and the $N$ communities,
$d\approx1$.

\begin{figure}
\center \includegraphics[width=7cm]{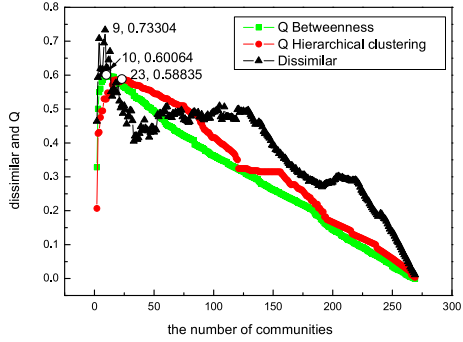}
\caption{The dissimilarity $D$ of the results from hierarchical
clustering and GN algorithm with same clusters.}\label{f}
\end{figure}

We use this algorithm to analyze the dissimilar of hierarchical
clustering and GN algorithm, and the result was shown in figure
\ref{f}. With the same number of communities, the $Q$ curves and
dissimilarity $D$ for the results from different algorithm are
shown. we also focus on the dissimilar of best division of them,
$D_{best}=0.756$, which means they are quite different.

\subsection{The influence of weight to the results of communities}
\label{analytic}

Now we turn to the effects of weight on the community structure of
weighted networks. In \cite{Li}, in order to study the impaction
of weight to the topological properties of network,we have
introduced the way to re-assign weights onto edge with $p=1,-1$
for weighted networks. Set $p=1$ represents the original weighted
network given by the ordered series of weights which gives the
relation between weight and edge but in a decreasing order,
\begin{equation}
W(p=1) = \left(w_{i_{1}j_{1}} = w^{1}\geq w_{i_{2}j_{2}} =
w^{2}\geq \cdots \geq
w_{\left(i_{L}\right)\left(j_{L}\right)}=w^{L}\right).
\end{equation}
$p=-1$ is defined as the inverse order as
\begin{equation}
W(p=-1) =\left(w_{i_{1}j_{1}} = w^{L}\leq \cdots \leq
w_{\left(i_{L-1}\right)\left(j_{L-1}\right)} = w^{2} \leq
w_{\left(i_{L}\right)\left(j_{L}\right)}=w^{1}\right),
\end{equation}
In this paper, we use the comparison of the communities which
formed in non-weighted and re-assign weights onto edges with
$p=1,-1$ to show the influence of weight to the results of
communities.

\begin{figure}
\center \includegraphics[width=7cm]{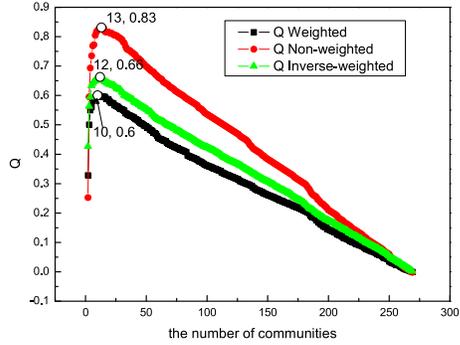}
\caption{the Q of different weights in GN algorithm}\label{d}
\end{figure}

\begin{figure}
\center
\includegraphics[width=7cm]{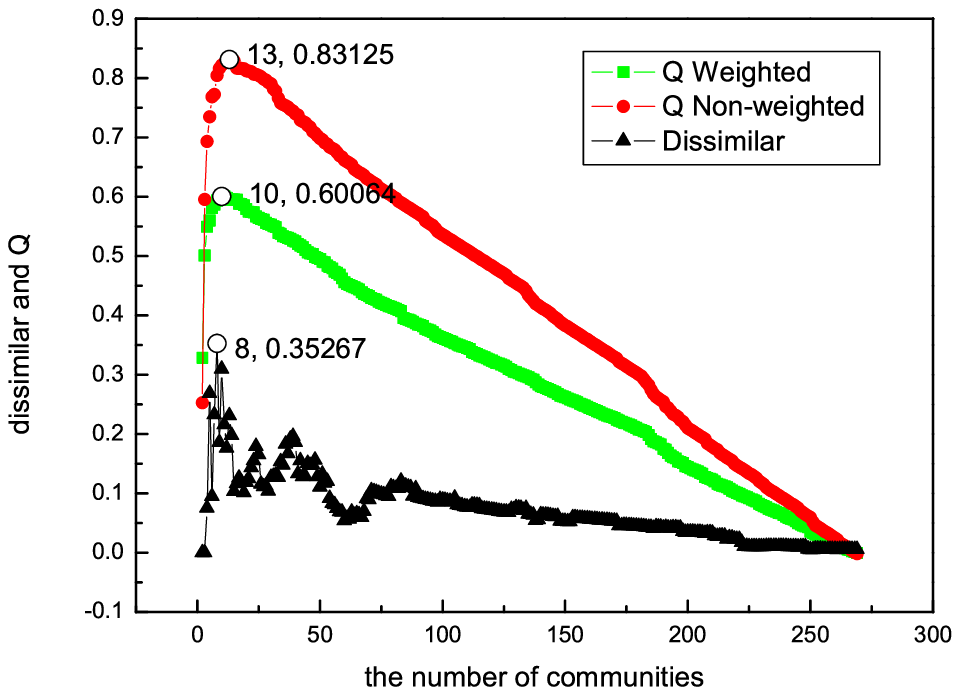}\includegraphics[width=7cm]{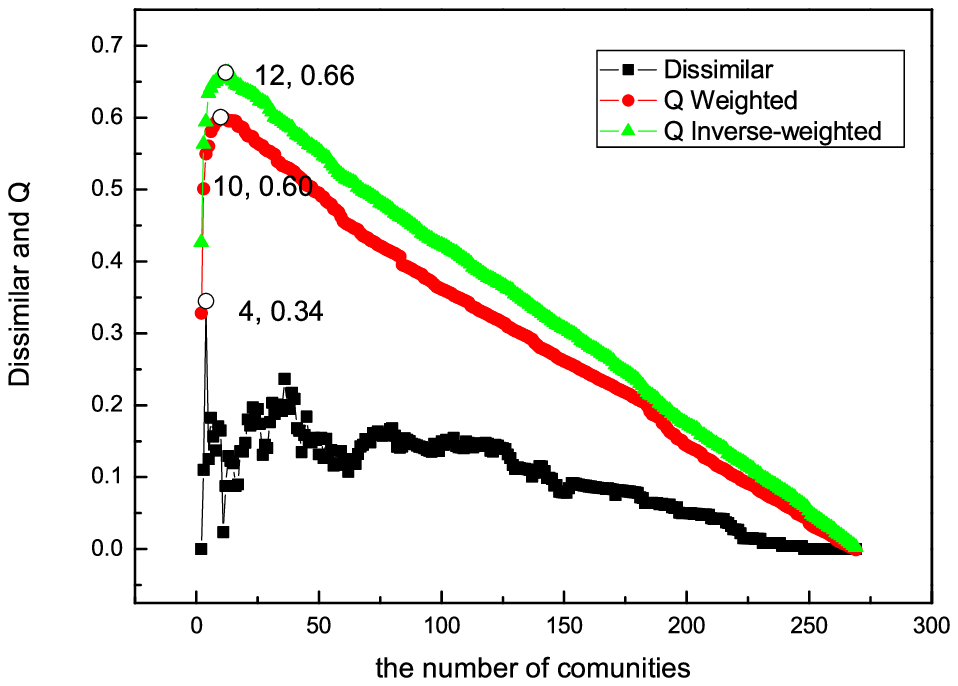}
\caption{A: The dissimilarity of non-weighted and weighted network
in GN algorithm. The dissimilarity of the best division is:
$D_{10,13}=0.42$.  B: The dissimilarity of weighted and inverse
weighted network in GN algorithm. The dissimilarity of the best
division is: $D_{10,12}=0.25$.}\label{e}
\end{figure}

We obtained the influence of weight to the results of communities
from the function Q and the dissimilarity. Using GN algorithm to
detect communities, the influence of weight were shown in figure
\ref{d},\ref{e}. In the figure \ref{d}, although the communities
number of best division in different weights are quite same, the
components of each community are quite different. The same things
happened in using hierarchical clustering to analyze the network.
Comparing these figures, we found that the weight have bigger
influence in GN algorithm.



\section{Concluding Remarks}\label{conclud}
In this paper, we study the community structure of scientists
collaboration network by using hierarchical clustering algorithm
and the algorithm of GN. And we also pay much attention to the
influence of the weight to results of communities. It has been
found that GN algorithm gives better results. Scientists who are
in the same university, institute or interested in similar
research topic are clustered to one community. In order to study
the topological role of the weight, we have introduced a measure
to describe the difference of two kinds of communities. Then we
investigate the different results of clustering for non-weighted,
weighted, and inverse weighted networks. The weight do have
influence on the formation of communities but it is not very
significant for our network of econophysicits. We guess that maybe
our network is a sparse network, so the existence or not of edges
have bigger influence to community structure of networks than the
weight.

\end{document}